\documentclass[12pt]{iopart}

\usepackage{graphicx}  
\begin{document}

\title[Hadronic dissipative effects on transverse dynamics at RHIC]{Hadronic dissipative effects on transverse dynamics at RHIC}

\author{T Hirano$^1$, U Heinz$^{2,3}$, D Kharzeev$^4$, R Lacey$^5$ and Y Nara$^6$}

\address{$^1$Department of Physics, The University of Tokyo, Tokyo 113-0033, Japan}

\address{$^2$Department of Physics, The Ohio State University, 
Columbus, OH 43210, USA}

\address{$^3$CERN, Physics Department, Theory Division, CH-1211 Geneva 23,
Switzerland}

\address{$^4$Nuclear Theory Group, Physics Department, Brookhaven 
            National Laboratory, Upton, NY 11973-5000, USA}

\address{$^5$Department of Chemistry, SUNY Stony Brook, Stony Brook, 
NY 11794-3400, USA}

\address{$^6$Akita International University, Akita 010-1211, Japan
}

\ead{hirano@phys.s.u-tokyo.ac.jp}

\begin{abstract}
We simulate the dynamics of Au+Au collisions at the Relativistic Heavy Ion 
Collider (RHIC) with a hybrid model that treats the quark-gluon 
plasma macroscopically as an ideal fluid, but models the 
hadron resonance gas microscopically using a hadronic cascade.
We find that
much of the mass-ordering pattern for $v_2(p_T)$
observed at RHIC is generated during 
the hadronic stage due to build-up of additional radial flow. 
We also find that the mass-ordering pattern is violated for $\phi$ meson 
due to small interaction cross section in the hadron resonance gas.
\end{abstract}


\section{Introduction}

Whether the quark-gluon plasma
(QGP)  behaves like a ``perfect liquid''
is one of the important question 
in heavy ion collisions at the Relativistic Heavy Ion
Collider (RHIC) \cite{experiments1,experiments2,experiments3,experiments4}.
The observed elliptic flow parameter $v_2$ and its transverse
momentum dependence
agree well with predictions from ideal 
fluid dynamics assuming zero viscosity \cite{reviews1, reviews2}.
The ideal fluid dynamical description, however,
gradually breaks down as one studies peripheral collisions
or moves away from midrapidity \cite{reviews1, reviews2}. 
This requires a more realistic treatment 
of the early and late stages
in dynamical modeling of relativistic 
heavy ion collisions.
In previous work 
\cite{HHKLN,HHKLN2} we have shown that a large
fraction of these deviations from ideal hydrodynamics is due to 
``late viscosity''.
Here we report additional results from the hybrid model study
focusing our attention on a detailed 
investigation of dissipative effects 
during the late hadronic rescattering 
stage.

\section{Model}

For the space-time evolution of the perfect QGP fluid we solve numerically
the equations of motion of ideal fluid dynamics for a given initial state
in three spatial dimensions and in time \cite{Hiranov2eta,HiranoTsuda}.
For the high temperature ($T>T_{\mathrm{c}}=170$ MeV) QGP phase we use the
equation of state of massless parton gas 
($u$, $d$, $s$ quarks and gluons) with a bag pressure $B$.
We switch from ideal hydrodynamics to 
a hadronic cascade model at the switching temperature 
$T_{\mathrm{sw}} = 169$\,MeV. The subsequent hadronic rescattering 
cascade is modeled by JAM \cite{jam}, initialized with hadrons 
distributed according to the hydrodynamic model output, calculated
with the Cooper-Frye formula \cite{CF} along the $T_{\mathrm{sw}}= 169$\,MeV 
hypersurface.
JAM implements experimental hadronic scattering cross section data where 
available and uses the additive quark model where data do not exist,
assuming the following formula 
for the total cross section \cite{jam,RQMD,RQMD2,UrQMD}:
\begin{eqnarray}
\label{eq:aqm}
  \sigma_{\mathrm{tot}} & = &
  \sigma_{NN}^{\mathrm{tot}}\frac{n_1}{3}\frac{n_2}{3} 
  \left(1-0.4\frac{n_{s1}}{n_1}\right)\left(1-0.4\frac{n_{s2}}{n_2}\right).
\end{eqnarray}
Here $\sigma_{NN}^{\mathrm{tot}}$ is the total nucleon-nucleon cross 
section, $n_{i}$ is the number of constituent quarks in a hadron and
$n_{si}$ is the number of strange quarks in a hadron. For hadrons 
composed entirely of strange quarks, such as $\phi$ = ($s\bar{s}$) and 
$\Omega$ = ($sss$), the cross sections become very small due to the 
suppression factors in brackets in Eq.~(\ref{eq:aqm}).
We note that, to study $\phi$ mesons in our hybrid model, we stabilize them by turning 
off their decay channels during the hadronic cascade.
For initial conditions in hydrodynamic equations, 
we employ the Glauber model suitably generalized 
to account for the longitudinal structure of particle multiplicity 
\cite{HHKLN,AG05}.

\section{Results}

%
 \begin{figure}[thb]
 \includegraphics[width=0.5\linewidth]{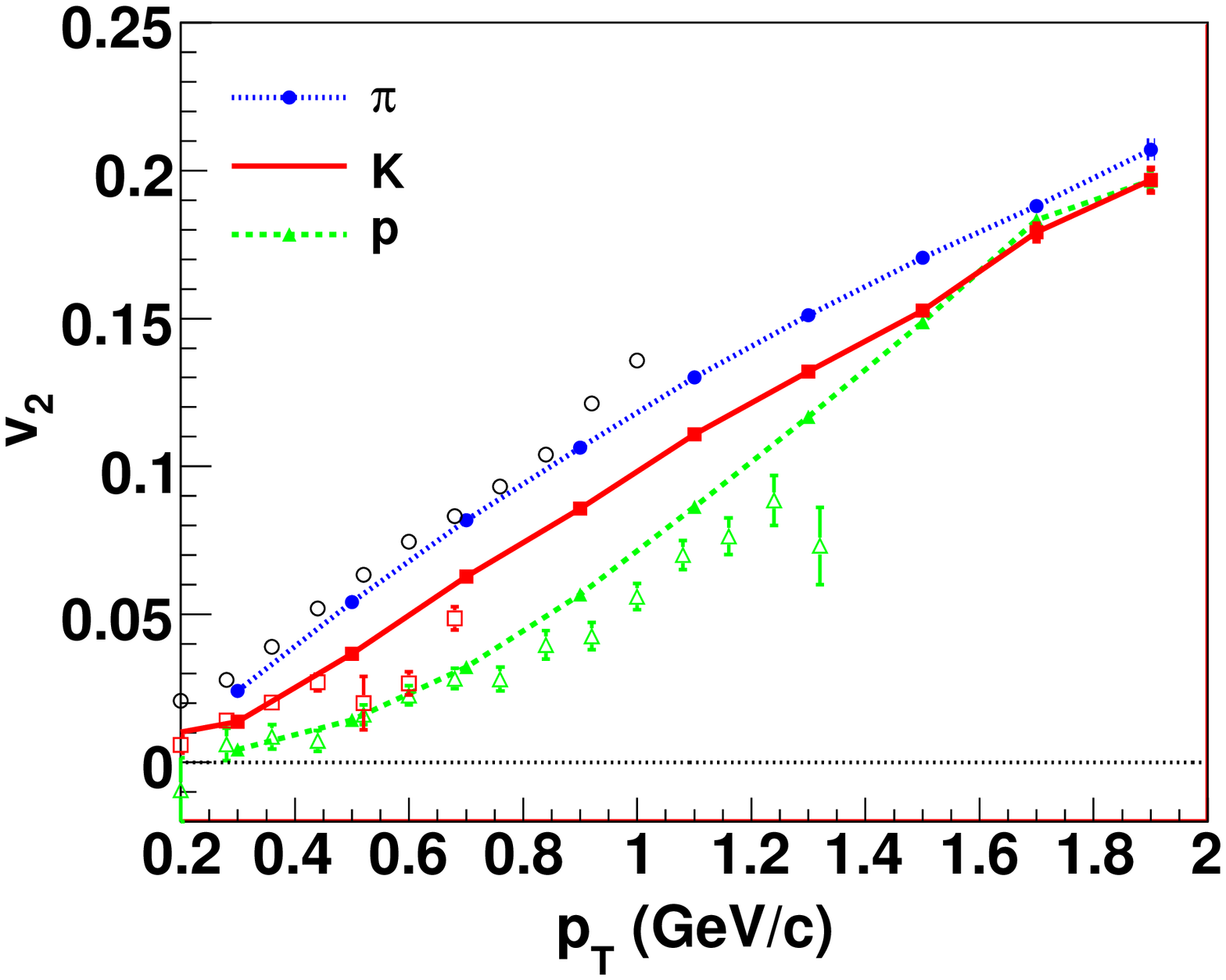}
 \includegraphics[width=0.5\linewidth]{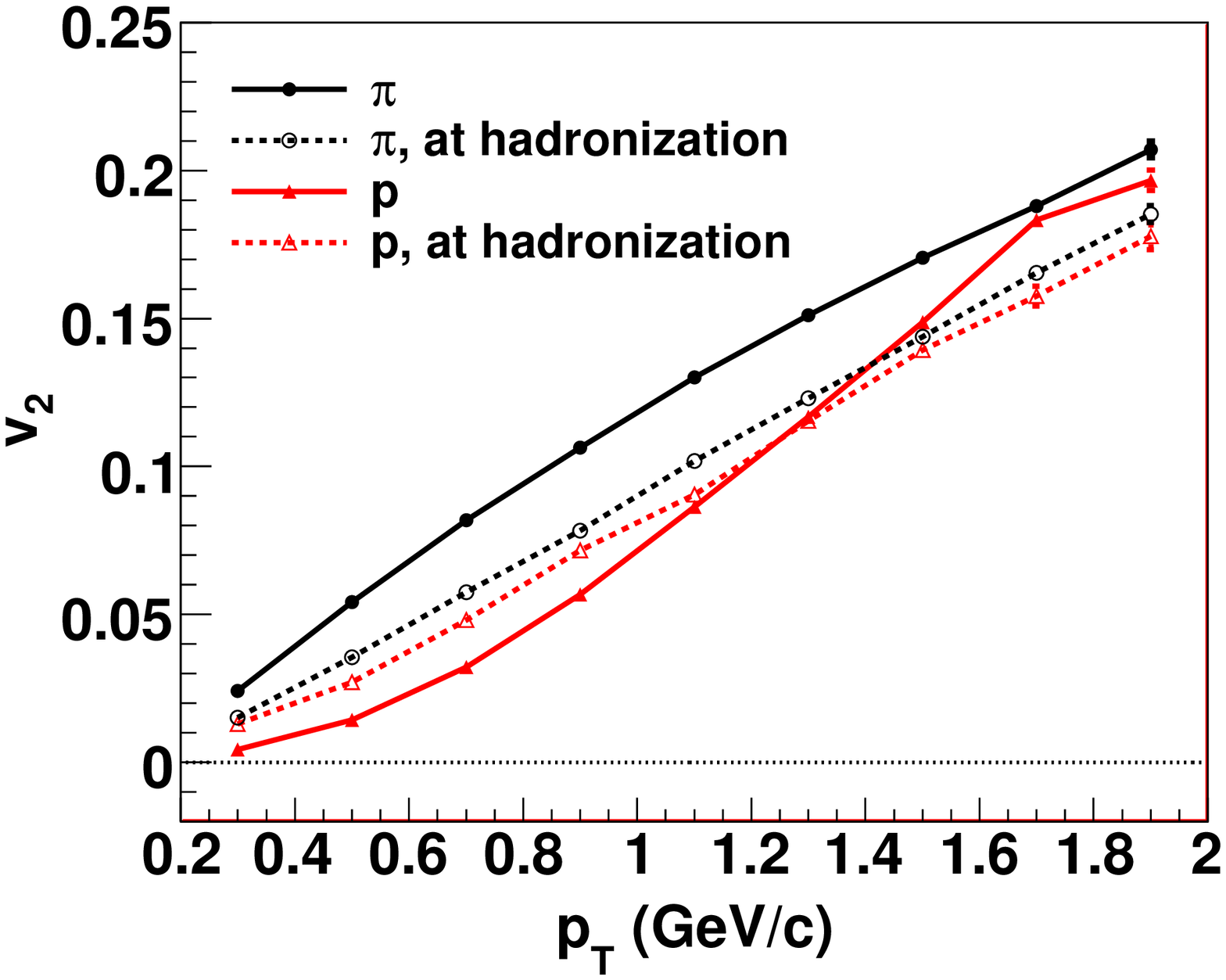}
 \caption{
  Transverse momentum dependence of the elliptic flow parameter
  $v_2(p_{\mathrm{T}})$. 
  (Left) $v_2(p_{\mathrm{T}})$ for pions, kaons and protons from the hybrid model
  are compared with STAR data \cite{star:anisotropy}.
  (Right) Solid (dashed) lines are with (without) 
  hadronic rescattering.
 }
 \label{fig:v2pt_before_after}
 \end{figure}
%

In Fig.~1 (left), we compare $v_2(p_{\mathrm{T}})$ for pions, kaons and protons
from the hybrid model with the STAR data \cite{star:anisotropy}.
We reasonably reproduce mass-splitting behaviour seen in the data.
We note that we also reproduce the data in other centralities (not shown)
except for very central collisions due to absence of eccentricity
fluctuation \cite{HHKLN3}.
Figure 1 (right) shows how mass-splitting of $v_2(p_{\mathrm{T}})$
is generated during evolution by switching off hadronic rescatterings
in a hadronic cascade.
Slope of pion $v_2(p_{\mathrm{T}})$ becomes steeper
due to additional development of elliptic flow and
reduction of mean $p_{\mathrm{T}}$ \cite{HG05}.
For heavy hadrons, on the other hand, radial flow reduces $v_2$ at low 
$p_{T}$ \cite{HKHRV01}.
Assuming positive elliptic flow, 
$v_\perp(\varphi{=}0,\pi) > v_\perp\left(\varphi{=}\frac{\pi}{2}, 
\frac{3\pi}{2}\right)$, the stronger transverse flow $v_\perp$ in the 
reaction plane pushes heavy particles to larger $p_{\mathrm{T}}$ more efficiently 
in the reaction plane than perpendicular to it.
The generation of additional radial flow \textit{in the 
hadronic stage} is responsible for the mass-splitting
of $v_2(p_{\mathrm{T}})$ observed in the low $p_{\mathrm{T}}$ region. 
From these observations we conclude that the large magnitude of the 
integrated $v_2$ and the strong mass ordering of 
$v_2(p_{\mathrm{T}})$ observed at RHIC result from a subtle interplay between 
perfect fluid dynamics of the early QGP stage and dissipative dynamics 
of the late hadronic stage.

%
 \begin{figure}[thb]
 \includegraphics[width=0.5\linewidth]{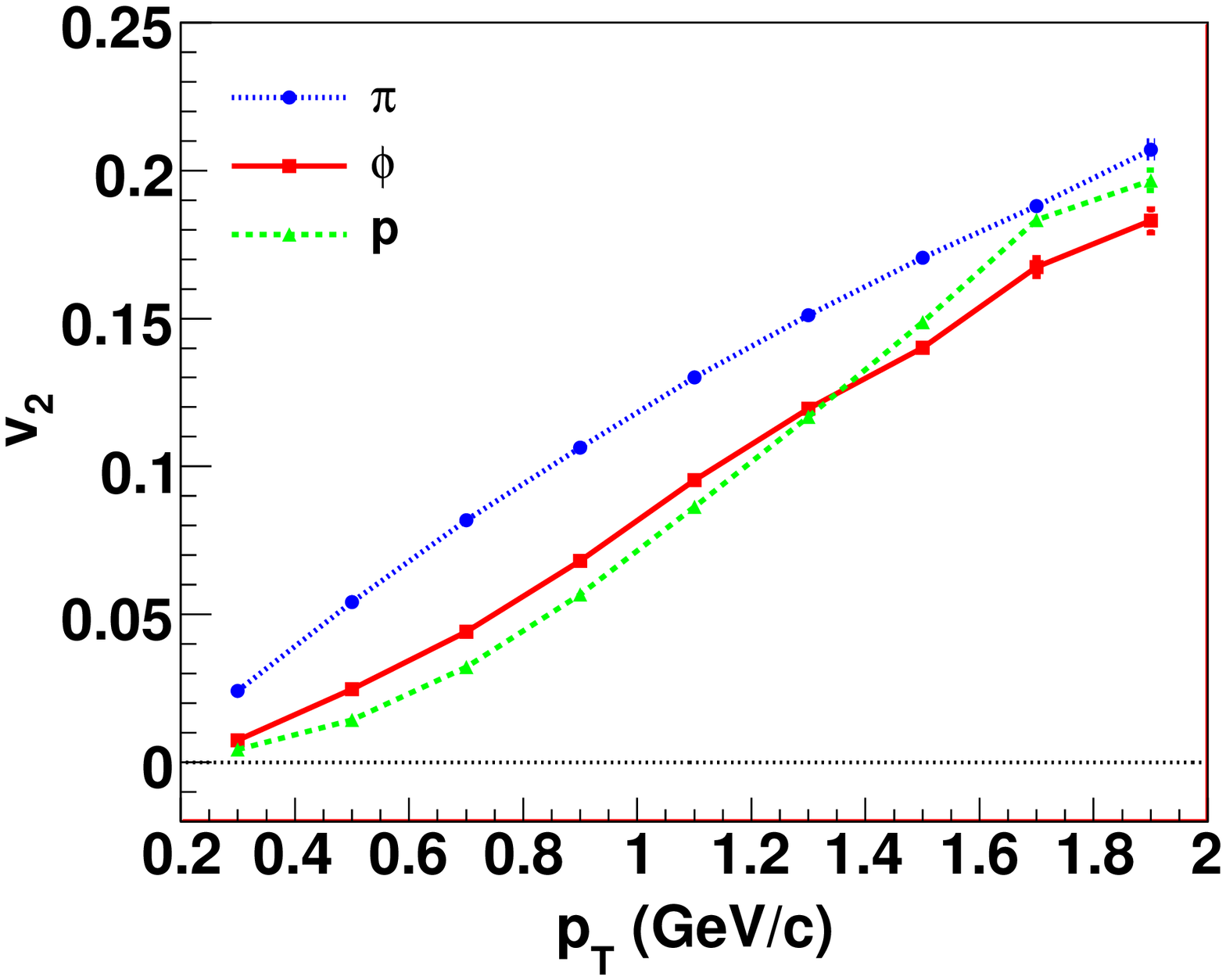}
 \includegraphics[width=0.5\linewidth]{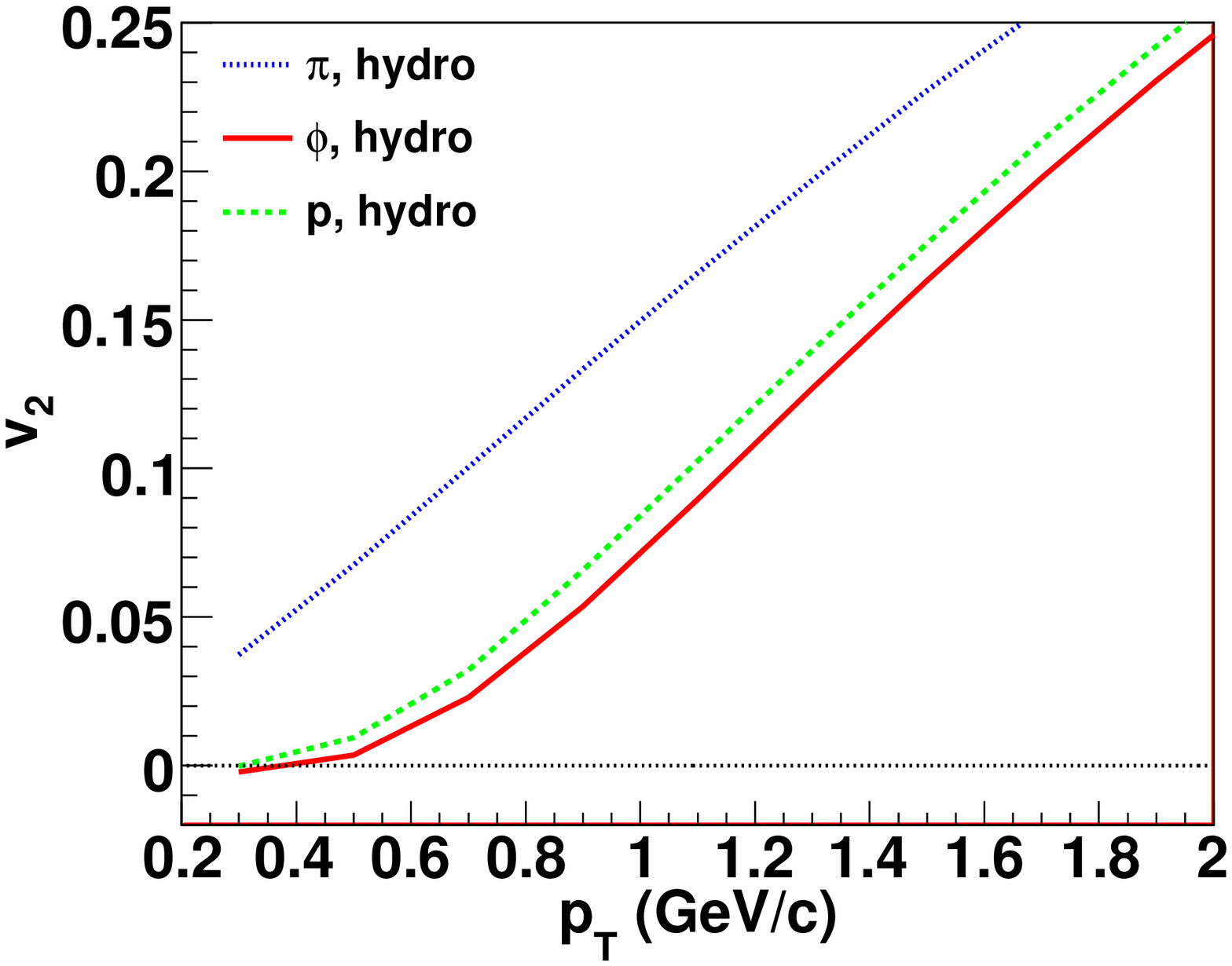}
 \caption{
(Left) $v_2(p_{\mathrm{T}})$ from the hybrid model.
(Right) $v_2(p_{\mathrm{T}})$ from ideal hydrodynamics with $T^{\mathrm{dec}}$=100 MeV.
Solid, dashed and dotted lines are results
for $\phi$ mesons, pions and protons.
 }
 \label{fig:v2pt}
 \end{figure}
%
$\phi$ mesons have considerably smaller scattering cross sections
than 
non-strange hadrons \cite{Shor:1984ui}. They are therefore expected to 
show larger dissipative effects in our hybrid model and not to fully 
participate in the additional radial flow generated during the hadronic 
rescattering stage.
Figure 2
shows $v_2(p_{\mathrm{T}})$ from the hybrid model (left)
and the ideal hydrodynamics (right) for 
$\pi$, $p$ and $\phi$.
As a result of rescattering the proton elliptic 
flow ends up being smaller than that of the $\phi$ meson, 
$v_2^p(p_T){\,<\,}v_2^\phi(p_T)$ for $0{\,<\,}p_T{\,<\,}1.2$\,GeV/$c$, 
even though $m_\phi{\,>\,}m_p$.
The large cross section difference 
between the protons and $\phi$ mesons in the hadronic rescattering 
phase leads to a violation of the hydrodynamic mass ordering at low 
$p_{\mathrm{T}}$ in the final state.

\section{Summary}

We have studied effects of hadronic dissipation on 
differential elliptic flow in Au+Au collisions at RHIC, using 
a hybrid model which treats the early QGP phase macroscopically as a 
perfect fluid and the late hadronic phase microscopically with a hadronic 
cascade.
The well-known mass-splitting of the differential 
elliptic flow observed in hydrodynamic models is seen to be 
mostly generated during the hadronic rescattering phase and to be largely 
due to a {\em redistribution} of the momentum anisotropy built up during 
the QGP stage. This redistribution is caused by the mass-dependent 
flattening of the transverse momentum spectra by additional radial flow 
generated during the hadronic stage. The much more weakly 
interacting $\phi$ mesons do not participate in this additional radial 
flow and thus are not affected by this redistribution of momentum 
anisotropies: their differential elliptic flow remains almost unaffected 
by hadronic rescattering. The net result of dissipative hadronic 
rescattering is therefore that the differential elliptic flow 
of protons {\em drops below} that of the $\phi$ mesons, in violation of 
the hydrodynamic mass-ordering.

\section*{Acknowledgments}

 This work was supported by the U.S. DOE under contracts 
 DE-FG02-01ER41190 (UH),
 DE-AC02-98CH10886 (DK)
 and DE-FG02-87ER40331.A008 (RL).
 The work of TH was partly supported by
 Grant-in-Aid for Scientific Research
 No.~19740130.

\end{document}